\documentclass[conference]{IEEEtran}
\usepackage{amsmath}
\usepackage{amsfonts}
\usepackage{epsfig}
\usepackage{amssymb}
\usepackage{graphics}
\usepackage{subfigure}

\begin{document}

\title{ \huge{Low-Complexity Adaptive Channel Estimation over Multipath Rayleigh Fading Non-Stationary Channels Under CFO}}

\author{Sayed A. Hadei, \emph{Student Member, IEEE,}
        and Paeiz Azmi, \emph{Senior Member, IEEE}\\Communication Systems Laboratory, Department of Electrical and Computer Engineering, \\Tarbiat Modares University, Tehran, Iran. P. O. Box: 14115-143\\  Email:
\{a.hadei, pazmi\}@modares.ac.ir}

\maketitle

\begin{abstract}

In this paper, we propose novel low-complexity adaptive channel
estimation techniques for mobile wireless channels in presence of
Rayleigh fading, carrier frequency offsets (CFO) and random
channel variations. We show that the selective partial update of
the estimated channel tap-weight vector offers a better trade-off
between the performance and computational complexity, compared to
the full update of the estimated channel tap-weight vector. We
evaluate the mean-square weight error of the proposed methods and
demonstrate the usefulness of its via simulation studies.
\end{abstract}

\IEEEpeerreviewmaketitle

\section{Introduction}
In wireless communications environment, signals suffer from
multiple reflections while travelling from the transmitter to the
receiver so that the receiver ends up getting several replicas of
the transmitted signal. The reflections are received with
different amplitude and phase distortions, and the overall
received signal is the combined sum of all the reflections. Based
on the relative phases of the reflections, the signals may add up
constructively at the receiver. Furthermore, if the transmitter is
moving with respect to the receiver, these destructive and
constructive interferences will vary with time. Therefore, the
performance of the wireless systems critically dependent on the
availability of accurate channel estimation \cite{proakis},
\cite{rappaport}, \cite{goldsmit}.

 Eke, periodic system variations resulting from  mismatches between
the transmitter and receiver carrier generators can be damaging to
the performance of the estimators, even for small carrier
frequency offsets \cite{Rupp}, \cite{Bahai}. Adaptive filtering
techniques are suitable for tracking of such time variant
channels. There are many adaptive filter algorithms that are
widely used for channel estimation, but they either have a high
mean-square error with slow convergence rate (such as, least mean
square (LMS) and normalized least mean square (NLMS) algorithms)
or a high computation complexity with fast convergence rate and
low mean-square error (such as, recursive least square (RLS)
algorithm) \cite{widrow}, \cite{haykin}, \cite{sayed}. Thus, mean-
square error (MSE), convergence rate and computational complexity
are three important points in selecting of adaptive algorithms for
channel estimation and this is considered in choice of the applied
algorithms. To address these problems, Ozeki and Omeda
\cite{ozeki} published the basic form of an affine projection
algorithm (APA) and Shin and Sayed developed tracking performance
analysis of a family of APA in \cite{sayed1}. Furthermore, in
order to reduce of computational complexity, one possible way is
that only part of the estimated tap-weight vector is selected for
update in each time iteration. In \cite{Aboulnasr}, \cite{Werner}
and \cite{Dogancay} adaptive filter algorithms with selective
partial update method (SPU) has been shown to have low
computational complexity and good performance. In this paper,
based on general formalism for the family of affine projection in
\cite{sayed1} and using the approaches that presented in
\cite{Aboulnasr} and \cite{Werner}, we present a novel algorithms
with general formalism which we called it low-complexity family of
affine projection algorithms based on selective partial update
method. Then, we use all of these algorithms for estimation of
mutipath Rayleigh fading non-stationary channels.

The rest of this paper is organized as follows. Section II
describe a brief introduction on the statistics of mobile wireless
channels. In Section III, we will have a breif review on some
known adaptive estimators and introduce cyclic non-stationary
channels model. Subsequently, the low-complexity family of affine
projection algorithms based on selective partial update method
will be presented in Section IV. Finally, Section V analyzes the
performance evaluation of the proposed estimation approaches by
computer simulation results to demonstrate the effectiveness of
the proposed algorithms for fading non-stationary channels
estimation in mobile wireless environments. Section VI concludes
this paper.
\section{Mobile Wireless Multipath Channel Model}
The complex baseband representation of the mobile wireless channel
impulse response can be described by \cite{steele}
\begin{equation}
\label{eq:1} h(t,
\tau)=\sum_{k=1}^{L}\gamma_{k}(t)x_{k}(\tau)\delta(\tau-\tau_{k})
\end{equation}
where $\tau_{k}$ and $\gamma_{k}(t)$ are respectively the delay
and path loss of the $k$-th path. $x_{k}(\tau)$ is a time-variant
complex fading sequence of the $k$-th path that models the
time-variations in the channel. Hence, the frequency response at
time $t$ is
\begin{equation}
\label{eq:2} H(t,f)\triangleq \int_{-\infty}^{+\infty}h(t,
\tau)e^{-j2\pi f \tau}d\tau
\end{equation}
without loss of generality, the sequence $x(\tau)$ is assumed to
have unit variance. Several mathematical models can be used to
characterize the fading properties of $x(\tau)$ and consequently
of the channel. A widely used model is known as Rayleigh fading.
In this case, for each $\tau$, the amplitude $|x(\tau)|$ is
assumed to have Rayleigh distribution \cite{goldsmit}. Therefore,
the sequence $\{x_{k}(\tau)\}$ are modeled as independent Rayleigh
fading sequences and the channel is referred to as a multipath
fading channel. In additional, the atu-correlation function of the
sequence $x(\tau)$, now regarded as a random process, is modeled
as a zeroth-order bessel function of the first kind, namely,
\begin{equation}
\label{eq:3} r(k)\triangleq E x(\tau)x(\tau-k)=J_{0}(2\pi
f_{D}T_{s}k).
\end{equation}
This commonly used choice of the auto-correlation function is
based on the assumption that all scatters are uniformly
distributed on a circle around the receiver, so that the power
spectrum of the channel fading gain $x(\tau)$, in continuous-time,
would have the following well-known $U$-shaped spectrum
\begin{equation}
\label{eq:4} S(f)=\frac{1}{\pi f_{D}\sqrt{1-\frac{f}{f_{D}}}},
~~~~~~~~~~~|f|\leq f_{D}
\end{equation}
In equation (\ref{eq:3}), $T_{s}$ is the sampling period of the
sequence $x(\tau)$, $f_{D}$ is called the maximum Doppler
frequency of the Rayleigh fading channel, and the function $J_{0}$
is defined by
\begin{equation}
\label{eq:5} J_{0}\triangleq \frac{1}{\pi}\int_{0}^{\pi}\cos(y\sin
\theta)d\theta.
\end{equation}
The Doppler frequency inversely proportional to the speed of
light, $c=3\times10^8 m/s$ and proportional to the speed of the
mobile user, $\nu$, and to the carrier frequency, $f_{c}$.

Due to the motion of the vehicle, $\gamma_{k}(t)$'s are modeled to
be wide-sense stationary (WSS), narrowband complex gaussian
process, which are independent for different paths. Furthermore,
$\gamma_{k}(t)$'s for all $k$ have the same normalized time
correlation function and different average powers
$\sigma_{k}^{2}$. In nonstationary channels, a first-order
approximation for the variation of a Rayleigh fading coefficient
$x(\tau)$ is to assume that $x(\tau)$ varies according to the
auto-regressive model
\begin{equation}
\label{ezafeh} x(\tau)=r(1)x(\tau-1)+\sqrt{1-|r(1)|^{2}}\eta(\tau)
\end{equation}
where $r(1)=J_{0}(2\pi f_{D}T_{s})$ and $\eta(\tau)$ denotes a
white noise process with unit-variance.

The two-ray \cite{winters}, typical urban (TU), and hilly terrain
(HT) \cite{steele}, \cite{ariyavisitakul} models are three
commonly used delay profiles. For the two-ray profile with equal
average power on each ray, the delay spread is $t_{d}/2$
($t_{d}=\tau_{1}-\tau_{0}$), i.e., a half of the delay different
between the two rays. The delay spreads for TU and HT delay
profiles are 1.06 and 5.04 $\mu$s, respectively \cite{geoffrey}.
\section{Overview on Some Known Adaptive Channel Estimators}
Considering Fig. 1, $ x(n)$ denotes a sequence that is transmitted
over an unknown channel of finite impulse response $w(n)$ of order
$M$. $d(n)$ and $e(n)$ are received signal and the output
estimation error, respectively. Consider the received signal
$d(n)$ assuming that arisen from the linear model
\begin{figure}[t]
\centering
\includegraphics[width=2.2in]{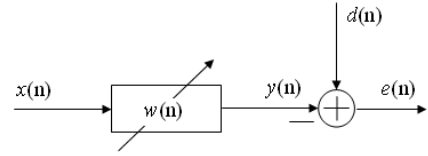}
\caption{A prototypical adaptive filter setup.} \label{figure1}
\end{figure}
 \begin{equation}
\label{eq:6} d(n)=u(n)w_{o}(n)+v(n)
\end{equation}
\begin{equation}
\label{A'} w_{o}(n)=w_{t}(n)e^{j\Psi n}
\end{equation}
where $w_{t}(n)$ is an unknown channel tap-weight vector and time
variant that we wish to estimate it, the term of $e^{j\Psi n}$
indicates the carrier frequency offsets \cite{Bahai}, $v(n)$ is
the measurement noise, assumed to be zero mean, white, gaussian
and independent of $u(n)$, and $u(n)=[x(n), x(n-1),..., x(n-M+1)]$
denotes $1\times M$ row input regressor vectors of the channel
with a positive-definite covariance matrix,
$R_{u}=E[u^{*}(n)u(n)]$.

In sequel, we shall adopt a model for the variation of $w_{t}(n)$.
Many models can be defined for this variation such as random-walk
or Markovian models \cite{Eweda}. One particular model that is
widely used in the adaptive filtering literature is a first order
random-walk model \cite{sayed1}, \cite{Lindbom}. On the other
hands, we know that the carrier frequency offsets between
transmitters and receivers and fading phenomenon are two important
trails of the mobile wireless channels. Therefore, we can use
achievable model for variations of fading channels until the
performance of adaptive estimators is enhanced.

For this reason, the following both cyclic and random
non-stationary model for variation of channel is introduced
\cite{Bahai}, \cite{sayed}
\begin{equation}
\label{A''} w_{t}(n)=w_{t}+\Xi(n)
\end{equation}
where $w_{t}$ is a constant vector that represent the non-fading
part of the channel, whereas $\Xi(n)$ represent the fading part of
channel that it can be modeled as autoregressive (AR) process of
order $\rho$ \cite{Giannakis}. We use common AR(1) model for this
process as
\begin{equation}
\label{A'''} \Xi(n+1)=\alpha\Xi(n)+q(n)
\end{equation}
where
\begin{align*}
0\leq\mid\alpha\mid<1
\end{align*}
and following from \cite{sayed}, the sequence $q(n)$ is assumed to
be i.i.d, zero-mean with autocorrelation matrix
$E\{q(n)q^{*}(n)\}=Q$ and $w_{t}(n)$ have a constant mean which we
shall denote by $w_{t}$, $Ew_{t}(n)\triangleq w_{t}$. In this
paper, $\alpha$ and autocorrelation matrix of $q(n)$ are $J_{0}(2
\pi f_{D} T_{s})$ and $Q=\sqrt{1-\alpha^{2}}I$, respectively. This
non-stationary model of channel is practical in wireless
communication systems and especially in channel estimation and
channel equalization applications.

 It is well known that the adaptive update scheme for
the LMS estimate of $w_{o}(n)$ is given by \cite{sayed}
 \begin{equation}
\label{eq:7} w(n+1)=w(n)+\mu u^{*}(n)e(n)
\end{equation}
where
\begin{align*}
e(n)=d(n)-u(n)w(n)
\end{align*}
is the output estimation error at time $n$ and $\mu$ is the step
size.

To increase the convergence speed of the LMS estimator, the NLMS
algorithm was proposed which can be state as \cite{sayed}
\begin{equation}
\label{eq:8} w(n+1)=w(n)+\frac{\mu}{\|u(n)\|^{2}}  u^{*}(n)e(n).
\end{equation}
For reducing the computational complicity, the estimated
tap-weight vector $w(n)$ and the excitation vector $u(n)$ of the
adaptive estimator can be partitioned into $B$ blocks with length
$L$ for each blocks that $B=M/L$ and it shall be an integer which
are defined as \cite{Dogancay}
\begin{equation}
\label{eq:9} u(n)=[x_{1}(n), x_{2}(n), ..., x_{B}(n)]
\end{equation}
\begin{equation}
\label{eq:10} w(n)=[w_{1}^{T}(n), w_{2}^{T}(n), ...,
w_{B}^{T}(n)]^{T}.
\end{equation}
The selective partial update NLMS algorithm for a single block
update at every iteration can be derived from the solution of the
following minimization problem \cite{Dogancay}
\begin{subequations}
\begin{align}
\label{eq:11} \min_{w_{j}(n+1)}\|w_{j}(n+1)-w_{j}(n)\|^{2}\\
subject~to~~  d(n)=u(n)w(n+1)
\end{align}
\end{subequations}
by using the method of Lagrange multipliers \cite{Goodwin}, the
update equation for selective partial update NLMS is given by
\begin{equation}
\label{eq:12}
w_{j}(n+1)=w_{j}(n)+\frac{\mu}{\|u_{j}\|^{2}}u_{j}^{*}(n)e(n)
\end{equation}
where \[j=\operatorname{arg\,max}\|u_{i}(n)\|^{2}\] for $
1\leqslant{i}\leqslant{B}$ and it denotes the number of blocks
that should be updated at each iteration.

From \cite{sayed1}, the general class of affine projection
algorithms can be stated as
\begin{align}
\label{eq:13} w(n+1)&=w(n-\beta(K-1))+\nonumber\\&\qquad{}\mu
U^{*}(n)(\epsilon I+U(n)U^{*}(n))^{-1}e(n)
\end{align}
where $e(n)=d(n)-U(n)w(n-\beta(K-1))$, and
\begin{subequations}
\begin{align}
\label{eq:14} U(n)=[u(n),u(n-D), ..., u(n-(K-1)D)]^{T}\\
d(n)=[d(n),d(n-D), ..., d(n-(K-1)D)]^{T}.
\end{align}
\end{subequations}
Based on (\ref{eq:13}) and by specific choices of the parameters
$\{K, D, \epsilon, \beta\}$ are resulted in different affine
projection algorithms. The particular choices and their
corresponding algorithms are summarized in Table \ref{table1}.
\begin{table}[!t]
\renewcommand{\arraystretch}{1.3}
\caption{APA FAMILY ALGORITHMS} \label{table1} \centering
\begin{tabular}{c|c|c|c|c}
\hline
\bfseries Algorithm & \bfseries $K$ & \bfseries $\epsilon$ & \bfseries $\beta$ & \bfseries $D$\\
\hline\hline NLMS&$K=1$&$\epsilon=0$&$\beta=0$&$D=1$\\
\hline
APA&$K\leq{M}$&$\epsilon=0$&$\beta=0$&$D=1$\\
\hline
BNDR-LMS&$K=2$&$\epsilon=0$&$\beta=0$&$D=1$\\
\hline
R-APA&$K\leq{M}$&$\epsilon\neq0$&$\beta=0$&$D=1$\\
\hline
PRA&$K\leq{M}$&$\epsilon\neq0$&$\beta=1$&$D=1$\\
\hline
NLMS-OCF&$K\leq{M}$&$\epsilon=0$&$\beta=0$&$D\geq1$\\
\hline
\end{tabular}
\end{table}
For NLMS-OCF, it is further assumed that $u(n-jD)$ is orthogonal
to $u(n), u(n-D), ..., u(n-(j-1)D)$ \cite{sayed1}. The motivation
for using $D>1$ is to increase the separation, and consequently
reduce the correlation. For PRA, the weight vector is updated once
every $K$ iterations that $K$ is positive integer and most
algorithms assume $K\leq{M}$.

\section{A Low-Complexity Family of Affine Projection
Algorithms} This section introduces a low-complexity family of
affine projection algorithms based on selective partial update
method. From \cite{aref}, the generic estimated tap-weight vector
update equation can be stated as
\begin{equation}
\label{eq:15} \boxed{w(n+1)=w(n-\beta(K-1))+\mu U^{*}(n)C(n)e(n)}
\end{equation}
where $C(n)$ is the $K\times{K}$ matrix and it is obtained from
Table \ref{table2}. This table shows that many classical and
modern adaptive filter algorithms can be derived through
(\ref{eq:15}).

 Considering (12), the constrained optimization
problem, which is solved by the proposed algorithms are given by
\begin{subequations}
\begin{align}
\label{eq:16} \min_{w_{h}(n+1)}\|w_{h}(n+1)-w_{h}(n)\|^{2}\\
subject~ to~~d(n)=U(n)w(n+1)
\end{align}
\end{subequations}
where $h=\{\tau_{1}, \tau_{2},...,\tau_{S}\}$ denotes the blocks
that updating at every adaptation. By using the lagrange
multipliers method and considering (\ref{eq:15}), the general
update equation of low-complexity family of affine projection
algorithms is given by
\begin{equation}
\label{eq:17} w_{h}(n+1)=w_{h}(n-K')+\mu U_{h}^{*}(n)C_{h}(n)e(n)
\end{equation}
where $C_{h}(n)$ is the $K\times{K}$ matrix and it is obtained
from Table \ref{table2}, $K'=\beta(K-1)$, and we have
\begin{equation}
\label{eq:18} U_{h}(n)=[U_{\tau_{1}}(n), U_{\tau_{2}}(n), ...,
U_{\tau_{S}}(n)]
\end{equation}
\begin{equation}
U_{\tau_{S}}(n)=[u_{\tau_{S}}(n),
u_{\tau_{S}}(n-D),...,u_{\tau_{S}}(n-(K-1)D)]^{T}
\end{equation}
\begin{equation}
\label{eq:19}U_{i}(n)=[u_{i}(n), u_{i}(n-D), ...,
u_{i}(n-(K-1)D)]^{T}.
\end{equation}
In above equations, $U_{h}(n)$ is the $K\times{LS}$ matrix,
$U_{\tau_{S}}(n)$ is $K\times{L}$, $u_{\tau_{S}}(n)$ is
$1\times{L}$ and $U_{i}(n)$ is the $K\times{L}$ matrix. For
obtaining the indices of $h$, we compute the values of
\begin{equation}
\label{eq:20}
\text{Tr}(U_{i}(n)U_{i}^{*}(n))~~~for~~~1\leq{i}\leq{B}
\end{equation}
then, $S$ largest blocks are selected for updating. The
  $K\times{K}$ matrix $U_{i}(n)U_{i}^{*}(n)$ is assumed to be full
  rank (i.e., invertable). The inequality $L\geq{K}$ is a necessary
  condition for $U_{i}(n)U_{i}^{*}(n)$ to be full rank.
Note that, for $K=1$, $\text{Tr}(U_{i}(n)U_{i}^{*}(n))$ reduces to
$\|u(n)\|^{2}$. Thus, (\ref{eq:20}) is consistent with the
selection criterion for the SPU-NLMS algorithm.

From (\ref{eq:17}), with different choices of the parameters $\{K,
D, \epsilon, \beta\}$ and $C_{h}(n)$ from Table.\ref{table2}, the
new adaptive estimators such as low-complexity version of
BNDR-LMS, NLMS-OCF and PRA based on selective partial update
method will be derived.
\begin{center}
\begin{table*}[!ht]
\renewcommand{\arraystretch}{2}
\caption{Correspondence Between Special Cases of (\ref{eq:17}) and
Various Low-Complexity Adaptive Filtering Algorithms}
\label{table2} \centering
\begin{tabular}{c|c|c|c|c|c}
\hline
\bfseries Algorithm & \bfseries $K$ & \bfseries $\beta$ & \bfseries $D$ & \bfseries $C(n)$ & \bfseries $C_{h}(n)$\\
\hline\hline LMS&$K=1$&$\beta=0$&$D=1$& $C(n)=1$& -\\
\hline NLMS&$K=1$&$\beta=0$&$D=1$& $C(n)=\|u(n)\|^{- 2}$&
$C_{h}(n)=[U_{h}(n)U_{h}^{*}(n)]^{-1}$\\
\hline
APA&$K\leq{M}$&$\beta=0$&$D=1$& $C(n)=[U(n)U^{*}(n)]^{-1}$& $C_{h}(n)=[U_{h}(n)U_{h}^{*}(n)]^{-1}$\\
\hline
BNDR-LMS&$K=2$&$\beta=0$&$D=1$& $C(n)=[U(n)U^{*}(n)]^{-1}$& $C_{h}(n)=[U_{h}(n)U_{h}^{*}(n)]^{-1}$\\
\hline
R-APA&$K\leq{M}$&$\beta=0$&$D=1$&$C(n)=[\epsilon{I}+U(n)U^{*}(n)]^{-1}$&$C_{h}(n)=[\epsilon{I}+U_{h}(n)U_{h}^{*}(n)]^{-1}$\\
\hline
PRA&$K\leq{M}$&$\beta=1$&$D=1$&$C(n)=[\epsilon{I}+U(n)U^{*}(n)]^{-1}$&$C_{h}(n)=[\epsilon{I}+U_{h}(n)U_{h}^{*}(n)]^{-1}$\\
\hline
NLMS-OCF&$K\leq{M}$&$\beta=0$&$D\geq1$&$C(n)=[U(n)U^{*}(n)]^{-1}$&$C_{h}(n)=[U_{h}(n)U_{h}^{*}(n)]^{-1}$\\
\hline
\end{tabular}
\end{table*}
\end{center}
\section{Performance Evaluation by Simulation Studeis}

Extensive computer simulations have been conducted to demonstrate
the performance of the adaptive estimators for multipath Rayleigh
fading non-stationary channel estimation. In adaptive filtering
problems, recursive algorithms are used to find the channel
tap-weight that produces the minimum error between the desired
value and received value. Since the channel can be modeled as a
tap-delay line filter, the channel estimation problem can be
formulated as that of an adaptive filtering problem. Fig.
\ref{figure1} illustrates the process of the adaptive filtering
and we can use of this process to estimate the non-stationary
channel taps. The objective of the adaptive algorithm is to find
the optimal tap-weight vector that gives the minimum mean-squared
error \cite{sayed}.
\begin{figure}[t!]
\centering
\includegraphics[width=3.5in]{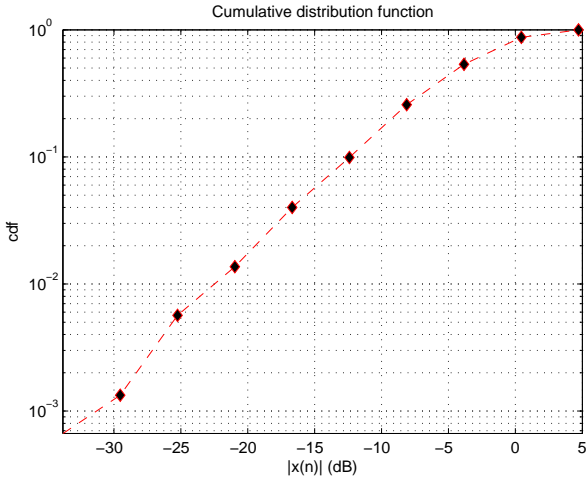}
\caption{The cumulative distribution function of a Rayleigh fading
sequence corresponding to a vehicle moving at 80Km/h.}
\label{figure2}
\end{figure}

\begin{figure}[t!]
\centering
\includegraphics[width=3.5in]{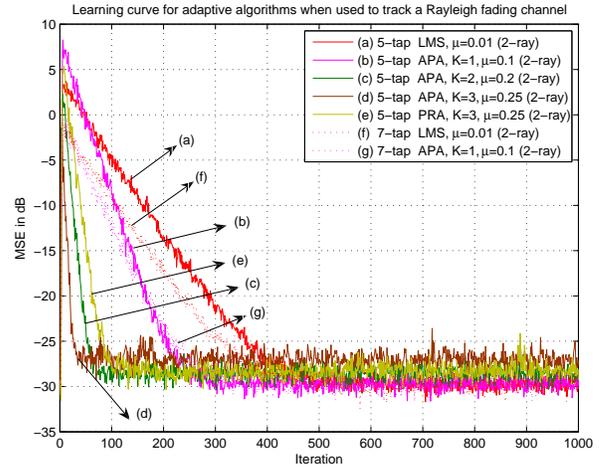}
\caption{The first 1000 samples of a learning curve that results
from tracking a Rayleigh fading non-stationary channel using LMS
and family of APA algorithms.} \label{figure3}
\end{figure}

\begin{figure}[t!]
\centering
\includegraphics[width=3.5in]{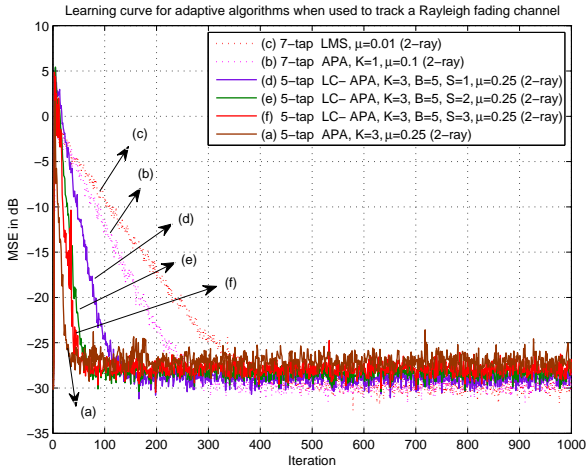}
\caption{The first 1000 samples of a learning curve that results
from tracking a Rayleigh fading non-stationary channel using LMS,
APA and Low-Complexity APA algorithms.} \label{figure4}
\end{figure}

\begin{figure}[t!]
\centering
\includegraphics[width=3.5in]{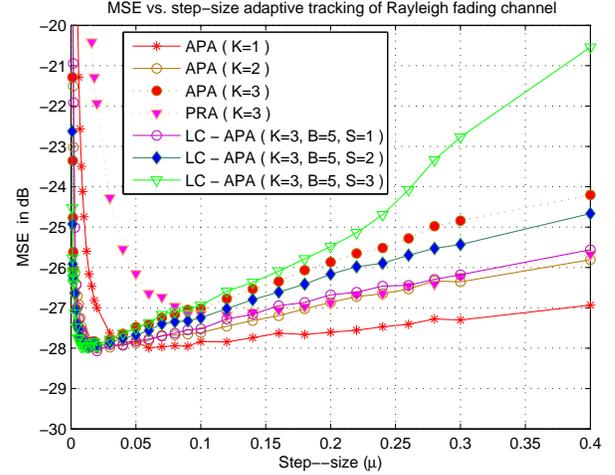}
\caption{A plot of the tracking MSE as a function of the step-size
for Rayleigh fading rays with Doppler frequency at 10Hz.}
\label{figure5}
\end{figure}

\begin{figure}[t!]
\centering
\includegraphics[width=3.5in]{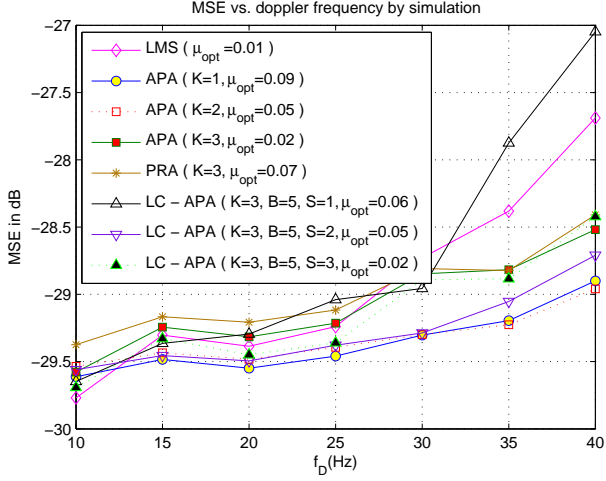}
 \caption{A plot of the tracking MSE as a function of the Doppler frequency
of the Rayleigh fading rays} \label{figure6}
\end{figure}

\begin{figure*}[!ht]
\centering \subfigure[]{
\includegraphics[width=1.4in, height=1.1in,]{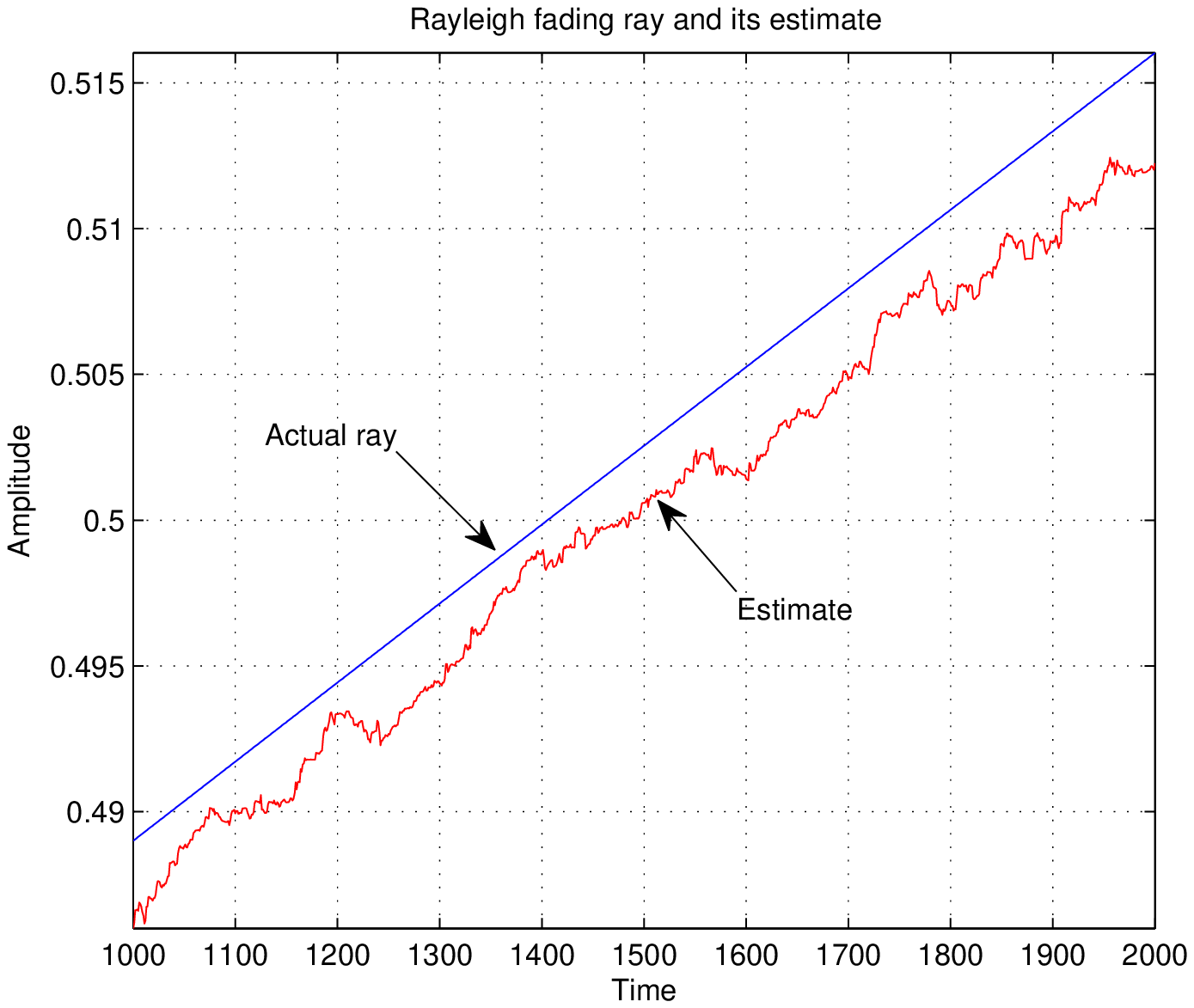}
\label{fig:subfig1} } \subfigure[]{
\includegraphics[width=1.4in, height=1.1in,]{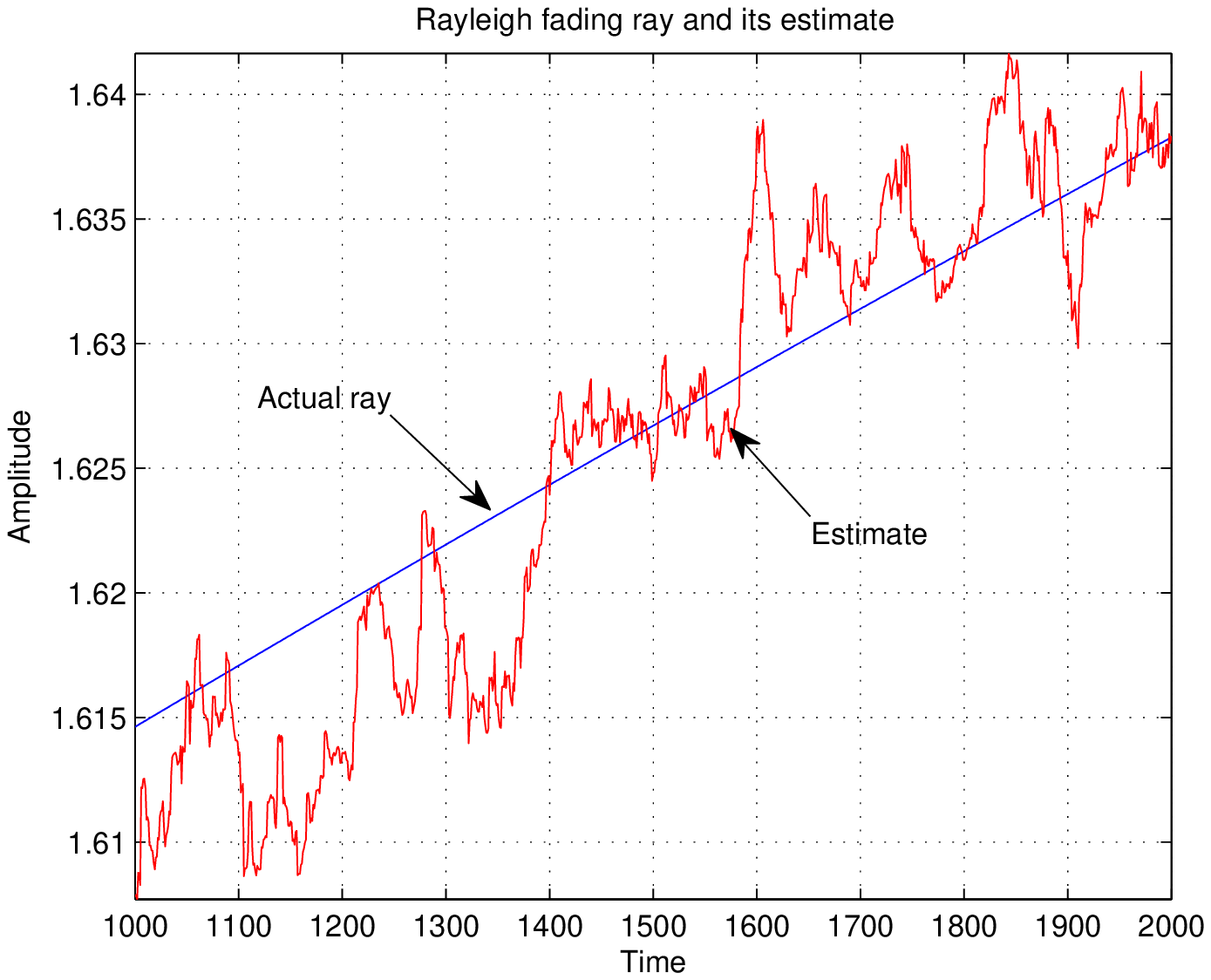}
\label{fig:subfig2} } \subfigure[]{
\includegraphics[width=1.4in, height=1.1in,]{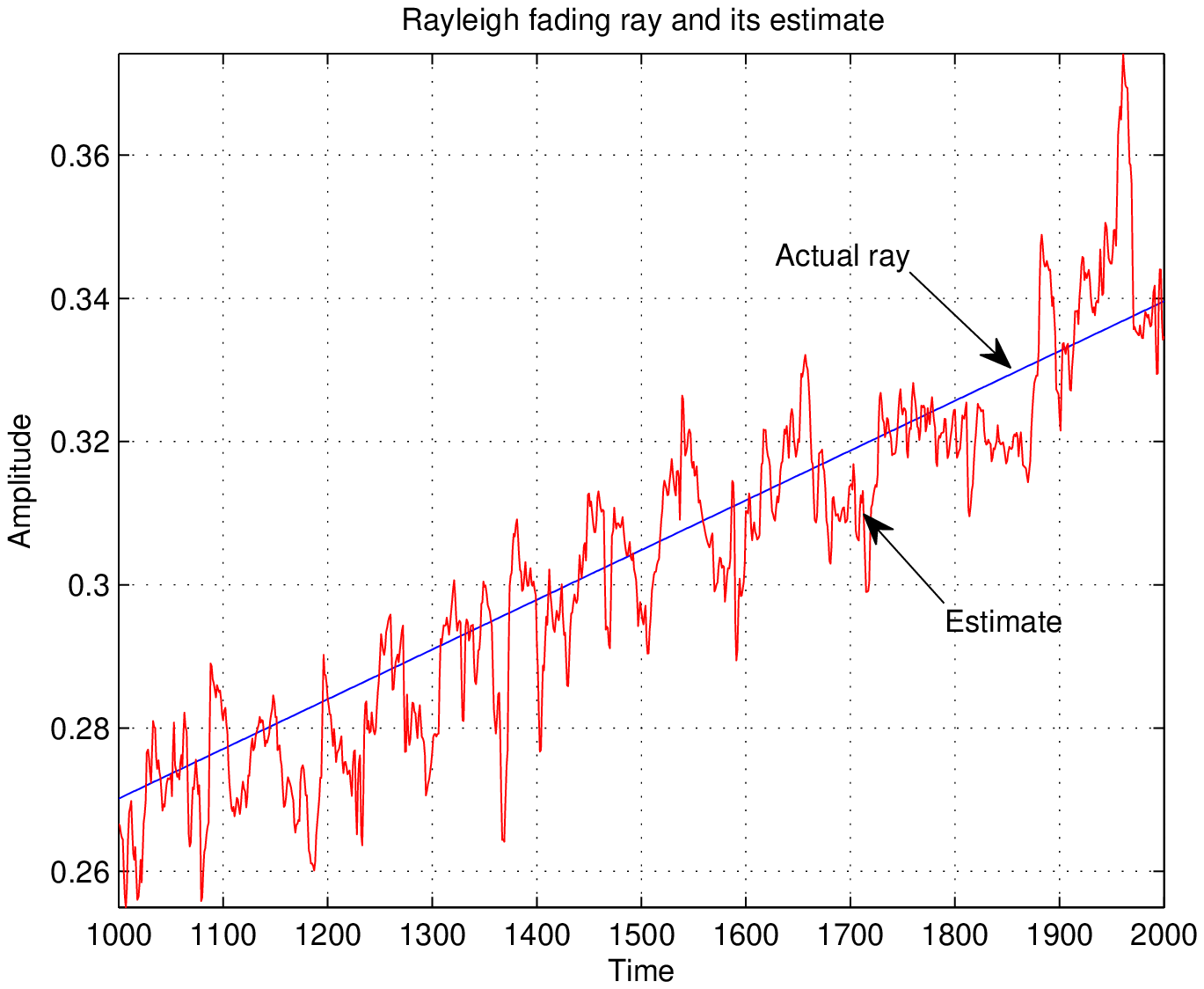}
\label{fig:subfig3} } \subfigure[]{
\includegraphics[width=1.4in, height=1.1in,]{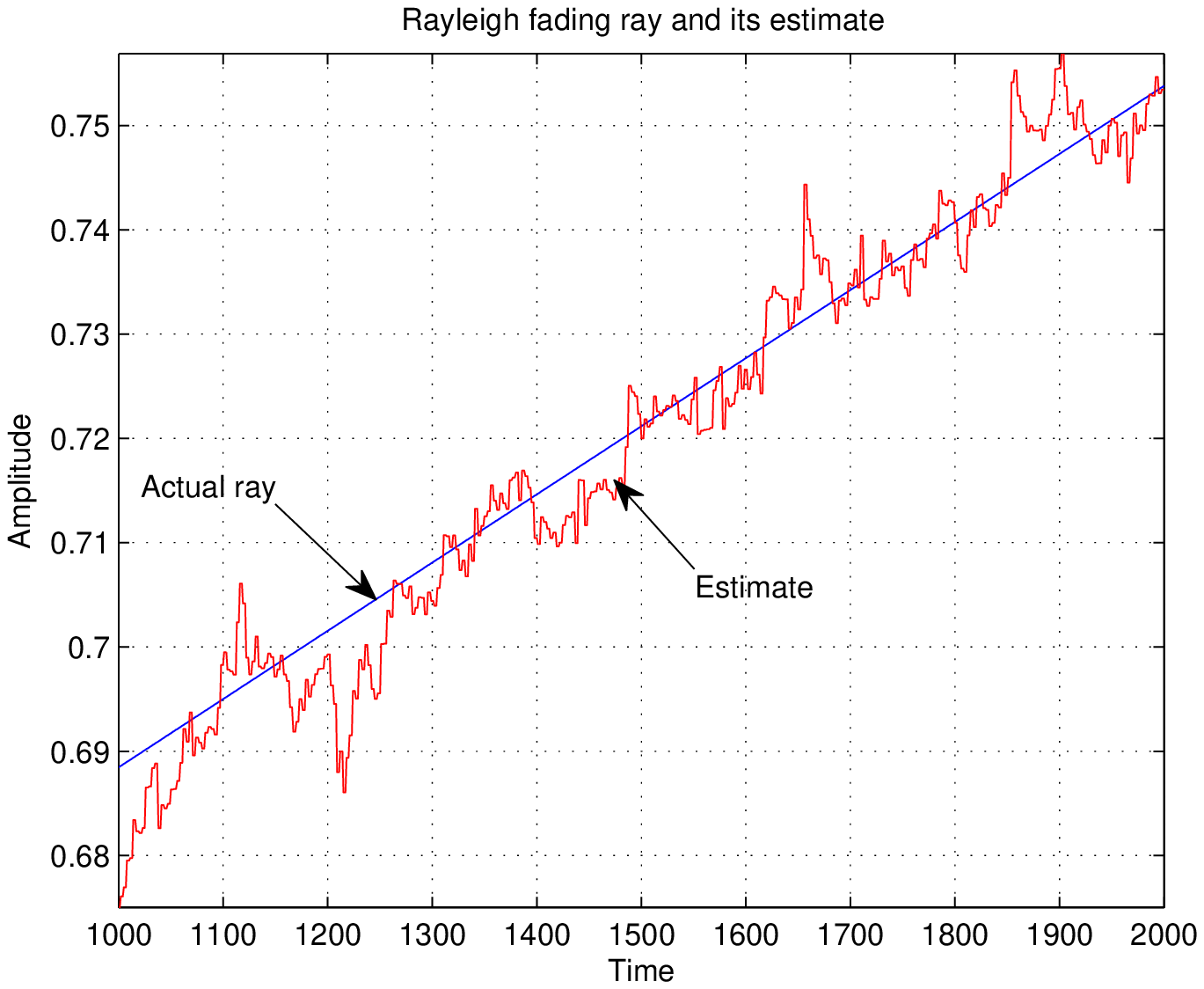}
\label{fig:subfig4}} \subfigure[]{
\includegraphics[width=1.4in, height=1.1in,]{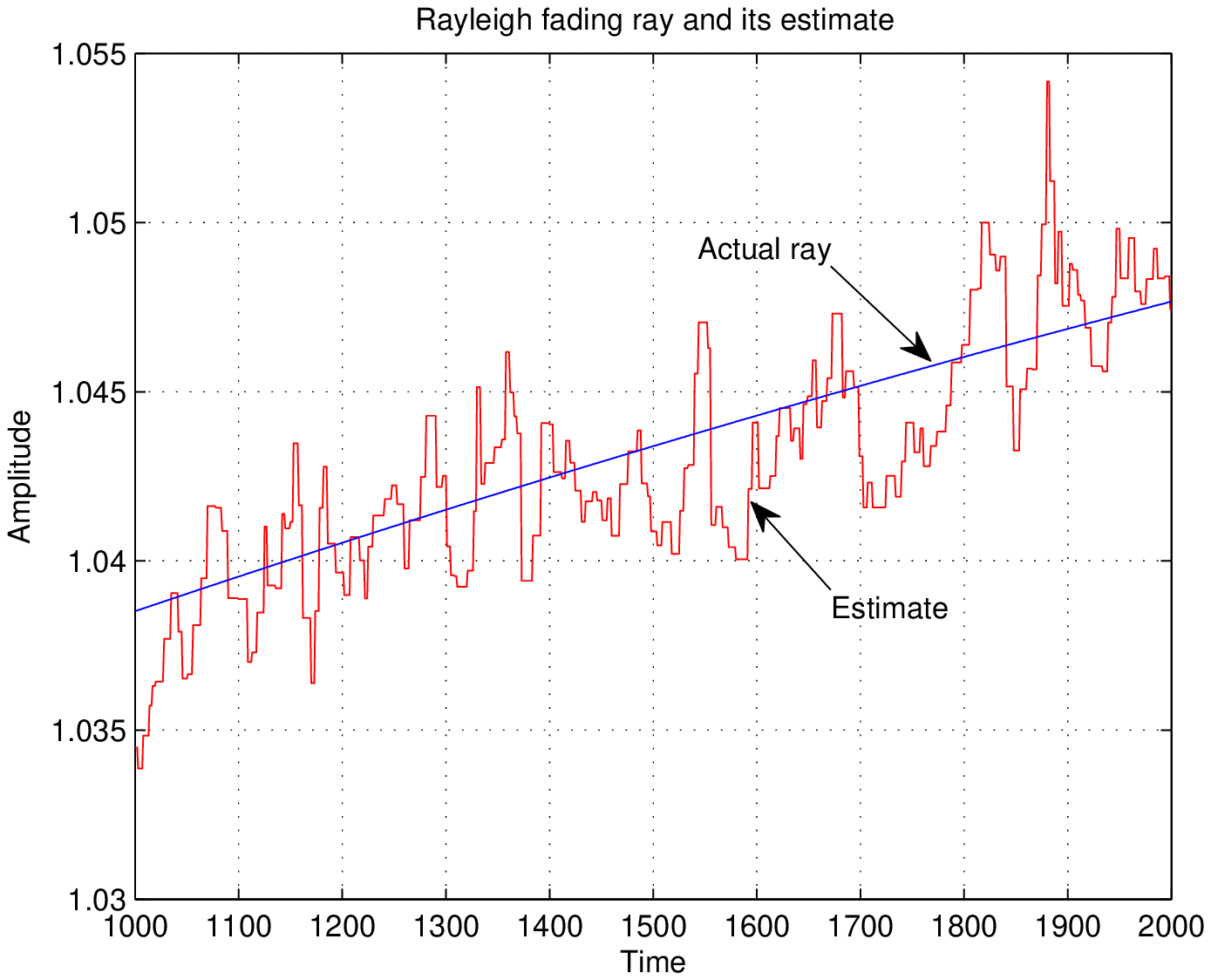}
\label{fig:subfig5}} \subfigure[]{
\includegraphics[width=1.4in, height=1.1in,]{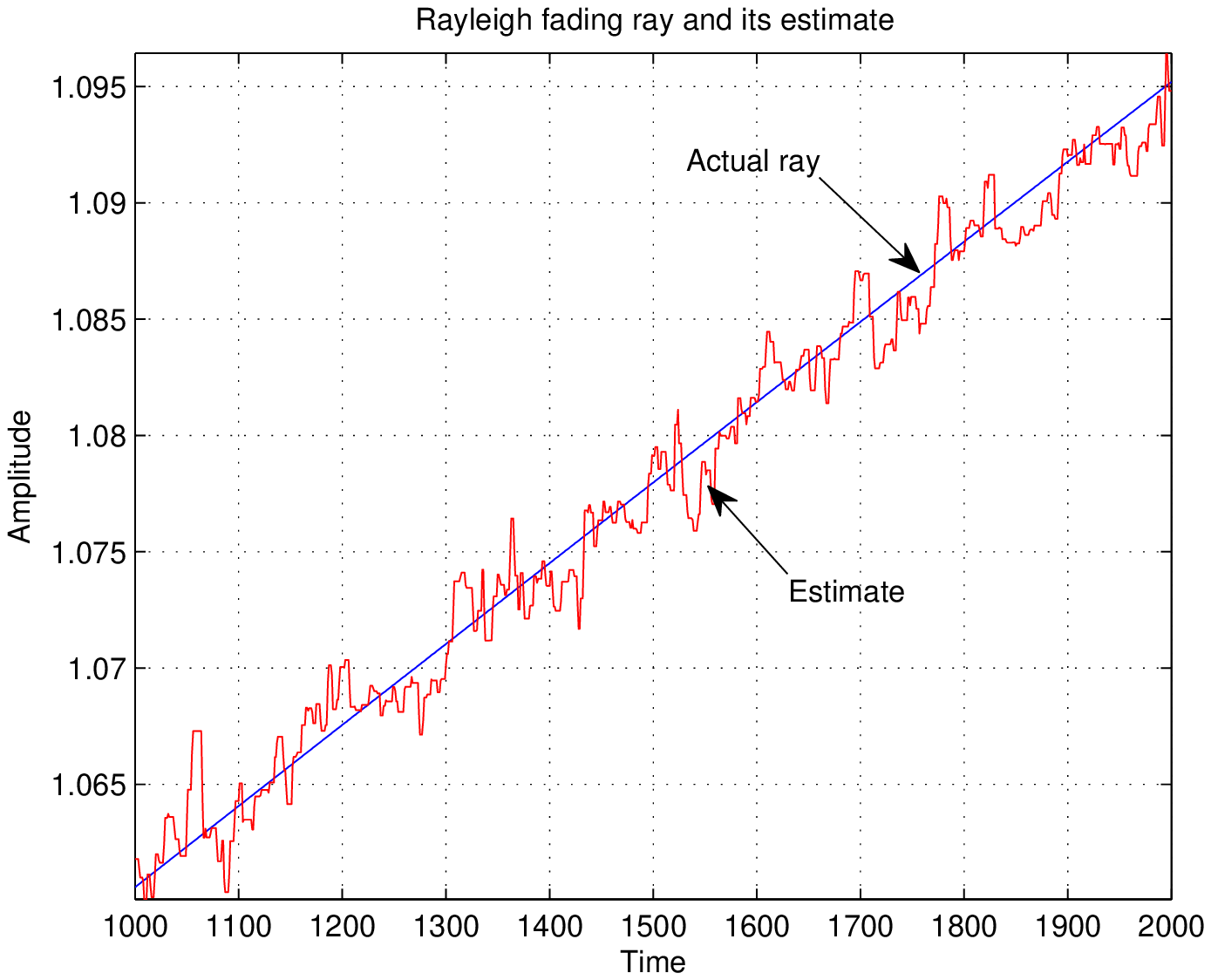}
\label{fig:subfig6}} \subfigure[]{
\includegraphics[width=1.4in, height=1.1in,]{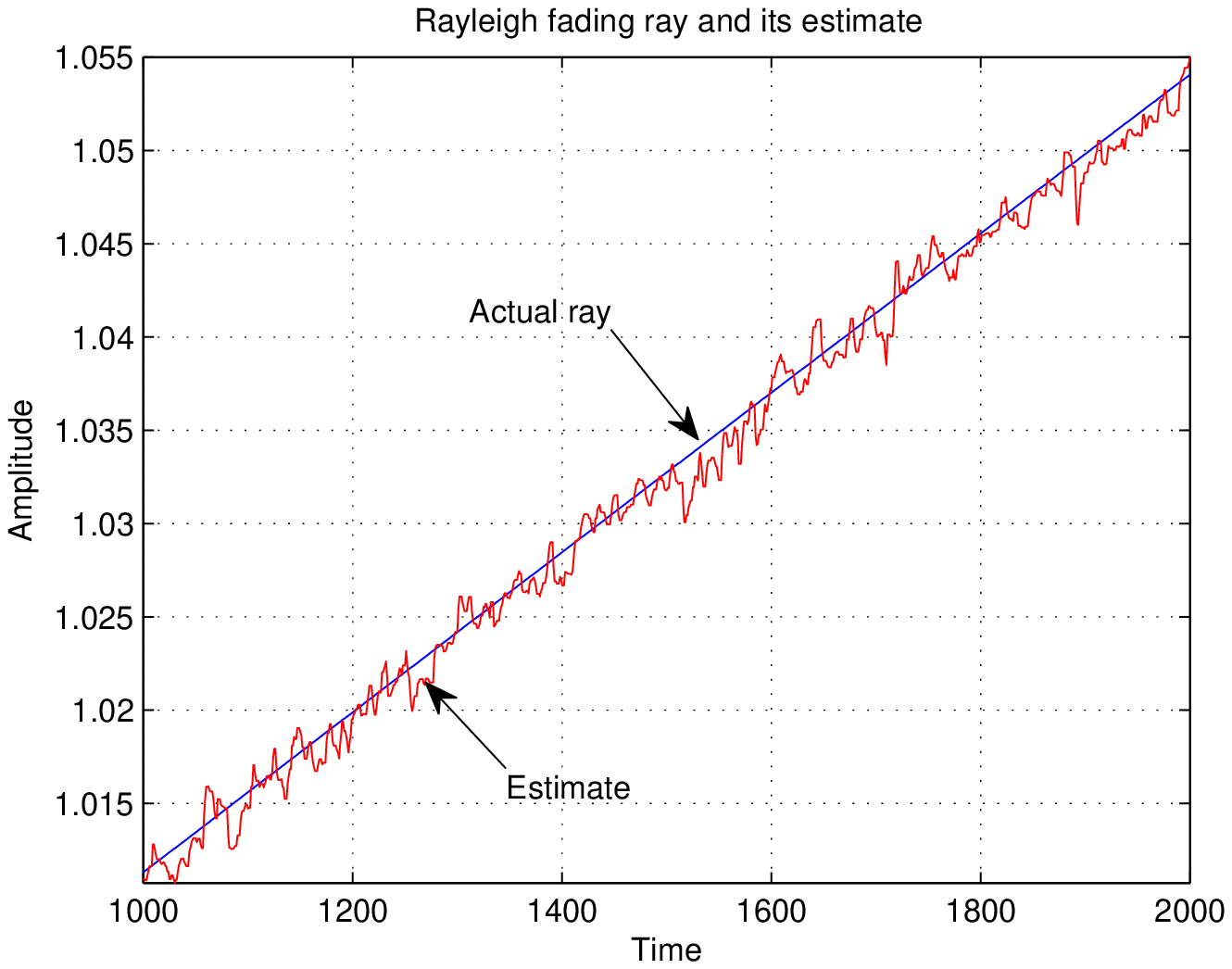}
\label{fig:subfig7}}

\label{figure7} \caption[Optional caption for list of figures]{A
plot of a typical trajectory of the amplitude of the first
Rayleigh ray and its estimate. The plots zooms on the interval
$1000\leq i \leq 2000$. \subref{fig:subfig1} {LMS algorithm with
$\mu=0.01$} \subref{fig:subfig2} {APA algorithm with $K=1$ and
$\mu=0.1$} \subref{fig:subfig3} {APA algorithm with $K=2$ and
$\mu=0.25$} \subref{fig:subfig4} {PRA algorithm with $K=3$ and
$\mu=0.25$} \subref{fig:subfig5} {LC-APA algorithm with $K=3$ ,
$B=5$ , $S=1$
 and $\mu=0.25$} \subref{fig:subfig6}
{LC-APA algorithm with $K=3$ , $B=5$ , $S=2$
 and $\mu=0.25$} \subref{fig:subfig7} {LC-APA algorithm with $K=3$ , $B=5$ , $S=3$
 and $\mu=0.25$}.}
\end{figure*}
In this paper, we consider a wireless channel with two Rayleigh
fading rays; both rays are assumed to fade at the same Doppler
frequency of $f_{D}=10Hz$. The channel impulse response sequence
consists of two zero initial samples (i.e., an initial delay of
two samples), followed by a Rayleigh fading ray, followed by
another zero sample, and by a second Rayleigh fading ray and then,
for another channel response, we will add two zero samples after a
second Rayleigh fading ray. In other words, we are assuming a
channel length of M=5 or 7 taps with only two active Rayleigh
fading ray, so that the weight vector that we wish to estimate has
the form

\begin{equation}
\label{eq:22} [~ 0~~ 0 ~~x_{2}(\tau)~~ 0~~ x_{4}(\tau)~]
\end{equation}
or

\begin{equation}
\label{eq:23} [ ~0~~ 0 ~~x_{2}(\tau) ~~0 ~~x_{4}(\tau) ~~0~~ 0~]
\end{equation}
Train adaptive algorithm to estimate and track these multipath
non-stationary channels. Assume random binary phase shift keying
(BPSK) input signal of unit variance is transmitted across the
channel and use it to excite the adaptive filter. Assume further
that the output of the channel is observed in the presence of
white additive Gaussian noise with variance $
\sigma_{\upsilon}^{2}=10^{-3}$, sampling period $T_{s}=0.8 \mu s$,
$\alpha=0.9$, $\sigma^{2}_{q}=10^{-4}$ and carrier frequency
offsets $\Psi=0.0001$. Fig. \ref{figure2} shows the cumulative
distribution function of the amplitude sequence. A plot of the
learning curve of all adaptive estimators is generated by running
them for 30000 iterations and averaging over 100 independent
experiments and show the good performance of proposed algorithm
with lower computational complexity that convergence rate of its
is comparable with full update APA estimators. Figs. \ref{figure3}
and \ref{figure4} show only the first 1000 iterations of a typical
learning curves. Fig. \ref{figure5} shows mean-square error (MSE)
curves versus step-size for the proposed scheme and other
algorithms that they are presented in this paper. In this
simulation, the Doppler frequency is fixed at 10Hz for both rays,
channel length $M=5$ taps, and the MSE values are obtained by
averaging over 100 independent realizations.

In Fig. \ref{figure6}, the Doppler frequency vary from 10Hz to
40Hz in increments of 5Hz and generate a plot of the MSE as a
function of the Doppler frequency. Run all adaptive estimators for
60000 iterations in each case and average the squared-error curve
over 1000 independent experiments. Also, we used the optimum value
of the step-size for all algorithms which we selected these
optimum values from minimizing of mean-square weight error
function over step-size $\mu$ \footnote{In Appendix A, we
calculate the mean-square weight error of the proposed method over
the cyclic non-stationary channel models. For the sake of arriving
to minimum mean-square error of estimation, we can minimize this
function over $\mu$.}. This figure shows the robustness of
adaptive estimators for different values of Doppler frequency.

Fig.7 shows typical trajectory of the amplitude of the first ray
and its estimate by different adaptive estimators. These plots
zooms onto the interval [1000, 2000]. The results show the good
tracking performance of adaptive algorithms for multipath Rayleigh
fading non-stationary channels tracking and proposed method is
better than LMS and APA ($K=1$) algorithms and comparable with APA
($K=3$) algorithm but our proposed method has lower computational
complexity than ordinary family of APA algorithms.

\section{Conclusion}
 The tracking performance of the new low-complexity
family of affine projection algorithms in the presence of carrier
frequency offsets and random channel nonstationarities was studied
in this paper. A general formalism for proposed algorithms was
also developed, and the mean-square weight error under
steady-state conditions was analyzed. The performance of our
methods were compared with those of some other estimators. The
proposed schemes presented a good trade-off between the
performance and the computational complexity. The simulation
results were included to demonstrate our claims.

\section*{Acknowledgment}
 This work was supported in part by the Iran Telecommunication Research Center
(ITRC) Tehran, Iran under grants 17581/500.

\appendices
\section{\small{The Mean-Square Weight Error of the Cyclic Non-Stationary Channel Estimation}}
To find the mean-square weight error of low-complexity family of
affine projection algorithms, following from \cite{sayed},
\cite{haykin}, let us praise a definition and a assumption as
follows

\emph{Definition}: For the cyclic model of channel
nonstationarities , the weight-error vector is
$\varepsilon(n)=w_{t}(n)e^{j\Psi n}-w_{h}(n)$, and a priori and a
posteriori estimation errors are defined as
$e_{a}(n)=U_{h}(n)\varepsilon(n-K')$,
$e_{p}(n)=U_{h}(n)(\varepsilon(n+1)-\varrho(n)e^{j\Psi n})$,
respectively. We also define the error-vector correlation matrix
and mean-square weight error as
$Z(n)=E[\varepsilon(n-K')\varepsilon^{*}(n-K')]$, $T(n)=Tr[Z(n)]$,
respectively.

\emph{Assumption}: The noise $v(n)$ is i.i.d and statistically
independent of the regression matrix $U_{h}(n)$ with variance
$\sigma^{2}_{v}=E|v(n)|^{2}$. Moreover, the random variables
$d(n)$, $v(n)$, $ U_{h}(n)$ have zero mean and
$U_{h}(n)U^{*}_{h}(n)$ is invertible and  $\varepsilon(n)$ is
statistically independent of $U_{h}(n)$. Furthermore,  The
sequence $q(n)$ is independent of the initial conditions
$\{w_{t}(-1), w(-1)\}$, of the $u(n), v(j)$ for all $n,j$ and of
the $d(j)$ for all $j<n$.

Thus, by using the above definition and  considering (\ref{A''})
and (\ref{A'''}), the weight-error vector recursion can be
rewritten as
\begin{equation}
\label{B}\varepsilon(n+1)=\varepsilon(n-K')+\varrho(n)e^{j\Psi
n}-\mu U_{h}^{*}(n)C_{h}(n)e(n)
\end{equation}
where
\begin{equation}
\label{B'}\varrho(n)=(w_{t}+\alpha\Xi(n)+q(n))e^{j\Psi}-(w_{t}+\Xi(n)).
\end{equation}
Introducing the noise vector as
\begin{align*}
 V^{T}(n)=[v(n)~~v(n-1)~~ \ldots
 ~~v(n-(K-1)D)].
\end{align*}
Then, from (\ref{eq:6}) and above assumption, the output
estimation error can be expressed in term of $e_{a}(n)$ as
\begin{equation}
\label{B''}e(n)=e_{a}(n)+V(n).
\end{equation}
In continues, considering above assumption and using (\ref{B''})
in (\ref{B}) and applying it for calculation of error-vector
correlation matrix $Z(n+1)$, we get
\begin{align*}
Z(n+1)&=E[\varepsilon(n+1)\varepsilon^{*}(n+1)]\\&=B_{1}(n)+B_{2}(n)+B_{3}(n)+B_{4}(n)
\end{align*}
where
\begin{align*}
&B_{1}(n)=Z(n)-\mu
Z(n)R_{N}+E[\varepsilon(n-K')\varrho^{*}(n)e^{-j\Psi
n}]\\&B_{2}(n)=E[\varrho(n)e^{j\Psi
n}\varepsilon^{*}(n-K')]+E\parallel\varrho(n)\parallel^{2}\\&\hspace{1.3cm}-\mu
E[\varrho(n)e^{j\Psi n}\varepsilon^{*}(n-K')]\\&B_{3}(n)=-\mu
R_{N}Z(n)+\mu^{2}Z(n)\Lambda(n)\\&\hspace{1.3cm}-\mu
R_{N}E[\varepsilon(n-K')\varrho^{*}(n)e^{-j\Psi
n}]\\&B_{4}(n)=\mu^{2}\sigma^{2}_{v}E[U_{h}^{*}(n)C_{h}(n)C_{h}^{*}(n)U_{h}(n)]
\end{align*}
and
\begin{align*}
&R_{N}=E[U_{h}^{*}(n)C_{h}(n)U_{h}(n)]\\\Lambda(n)=&E[U_{h}^{*}(n)C_{h}(n)U_{h}(n)U_{h}^{*}(n)C_{h}^{*}(n)U_{h}(n)].
\end{align*}
For the sake of obtain the error-vector correlation matrix, we
require to evaluate the cross correlation between the tap-weight
error vector and $\varrho(n)$, and
$E\parallel\varrho(n)\parallel^{2}$. At first, by taking the
squared norm and expectation from both side of (\ref{B'}), we
obtain
\begin{align}
\label{B'''} E[\|\varrho(n)\|^{2}]=&Tr(Q)+\mid 1-e^{j\Psi}
\mid^{2} Tr(W_{t})\nonumber\\&+\mid 1-\alpha e^{j\Psi}
\mid^{2}Tr(\Theta)
\end{align}
where
\begin{align*}
W_{t}=w_{t}w_{t}^{*}~~~~~~~~~ \Theta=\mathop {\lim }\limits_{n \to
\infty } E(\Xi(n)\Xi^{*}(n)).
\end{align*}
From (\ref{A''}) and (\ref{A'''}) we can find that
\begin{align*}
\Theta=\frac{Q}{1-|\alpha|^{2}}.
\end{align*}
Then, in steady-state conditions, following the derivation in
\cite{Yousef}, we can define that
\begin{align*}
&E(\varepsilon(n-K'))=H(n)=He^{j\Psi
n}\\E(&\varepsilon(n-K')\Xi^{*}(n))=Y(n)=Ye^{j\Psi n}
\end{align*}
where
\begin{align*}
&H=[I-\mu
R_{N}-e^{j\Psi}I]^{-1}w_{t}(1-e^{j\Psi})\\Y=[&\alpha^{*}(I-\mu
R_{N})-e^{j\Psi}I]^{-1}(\alpha^{*}(1-\alpha
e^{j\Psi})\Theta-e^{j\Psi}Q).
\end{align*}
Therefore, using (\ref{B'}) and above expressions, we can expand
$B_{1}(n)$, $B_{2}(n)$ and $B_{3}(n)$ as following
\begin{align*}
&B_{1}(n)=Z(n)-\mu
Z(n)R_{N}+(e^{-j\Psi}-1)w_{t}^{*}H\\&\hspace{1.3cm}+(\alpha^{*}e^{-j\Psi}-1)Y\\&B_{2}(n)=(e^{j\Psi}-1)w_{t}H^{*}+(\alpha
e^{j\Psi}-1)Y^{*}+Tr(Q)\\&\hspace{1.3cm}+\mid 1-e^{j\Psi} \mid^{2}
Tr(W_{t})+\mid 1-\alpha e^{j\Psi}
\mid^{2}Tr(\Theta)\\&\hspace{1.3cm}-\mu
R_{N}(e^{j\Psi}-1)w_{t}H^{*}-\mu R_{N}(\alpha
e^{j\Psi}-1)Y^{*}\\&B_{3}(n)=-\mu
R_{N}Z(n)+\mu^{2}Z(n)\Lambda(n)\\&\hspace{1.3cm}-\mu
R_{N}(e^{-j\Psi}-1)w_{t}^{*}H-\mu R_{N}(\alpha^{*}e^{-j\Psi}-1)Y.
\end{align*}
Thus, by calculating error-vector correlation matrix and applying
the definition of $T(.)$ in it, we obtain
\begin{align*}
T(n+1)=&(1-2\mu
Tr(R_{N})+\mu^{2}Tr(\Lambda(n)))T(n)\\&+\mu^{2}\sigma^{2}_{v}Tr(E[U_{h}^{*}(n)C_{h}(n)C_{h}^{*}(n)U_{h}(n)])+F.
\end{align*}
Finally, using steady-state conditions $T(\mu)=\mathop {\lim
}\limits_{n \to \infty }T(n,\mu)=\mathop {\lim }\limits_{n \to
\infty }T(n+1,\mu)$, mean-square weight error of cyclic
non-stationary channel estimation may be given as
\begin{align*}
T(\mu)=&\frac{\mu^{2}\sigma^{2}_{v}Tr(E[U_{h}^{*}(n)C_{h}(n)C_{h}^{*}(n)U_{h}(n)])+F}{2\mu
Tr(R_{N})-\mu^{2}Tr(\Lambda(n))}
\end{align*}
where
\begin{align*}
F=&|1-e^{j\Psi}|^{2}Re\{Tr(W_{t}(I-2F_{\alpha}))\}\\&+|1-\alpha
e^{j\Psi}|^{2}Re\{Tr(\Theta(I-2\alpha^{*}F_{\beta}))\}\\&+Re\{Tr(Q(I-2(\alpha^{*}-e^{j\Psi})F_{\beta}))\}
\end{align*}
and
\begin{align*}
&F_{\alpha}=(I-\mu R_{N})[I-\mu
R_{N}-e^{j\Psi}I]^{-1}\\F&_{\beta}=(I-\mu R_{N})[\alpha^{*}(I-\mu
R_{N})-e^{j\Psi}I]^{-1}
\end{align*}

\bibliographystyle{IEEEtran}
\bibliography{IEEEabrv,mybibfile1}
\end{document}